\documentclass[preprint,aps,prb,epsfig]{revtex4}

\usepackage{graphics}
\usepackage{epsfig}

\begin{document}


\title{Stochastic Dynamical Structure (SDS) of Nonequilibrium Processes in the Absence of Detailed Balance. II:  \\ construction of SDS with nonlinear force and multiplicative noise }
\author{  P. Ao }
\address{Departments of Mechanical Engineering and Physics, 
               University of Washington, Seattle, WA 98195, USA }
\date{March 30, 2008 }


\begin{abstract}
There is a whole range of emergent phenomena in non-equilibrium behaviors can be well described by a set of stochastic differential equations. Inspired by an insight gained during our study of robustness and stability in phage lambda genetic switch in modern biology, we found that there exists a classification of generic nonequilibrium processes: In the continuous description in terms of stochastic differential equations, there exists four dynamical elements: the potential function $\phi$, the friction matrix $ S$ , the anti-symmetric matrix $ T $, and the noise. The generic feature of absence of detailed balance is then precisely represented by $T$.
For dynamical near a fixed point, whether or not it is stable or not, the stochastic dynamics is linear. A rather complete analysis has been carried out (Kwon, Ao, Thouless, cond-mat/0506280; PNAS, {\bf 102} (2005) 13029), referred to as SDS I.        
One important and persistent question is the existence of a potential function with nonlinear force and with multiplicative noise, with both nice local dynamical and global steady state properties.  Here we demonstrate that a dynamical structure built into stochastic differential equation allows us to construct such a global optimization potential function. First, we provide the construction. One of most important ingredient is the generalized Einstein relation. We then present an approximation scheme: The gradient expansion which turns every order into linear matrix equations. 
The consistent of such methodology with other known stochastic treatments will be discussed in next paper, SDS III; and the explicitly connection to statistical mechanics and thermodynamics will be discussed in a forthcoming paper, SDS IV. 
(The main results were published. Please cite the present paper as {\it Potential in Stochastic Differential Equations: Novel Construction}, P. Ao, J. Phys. {\bf A37} L25-L30 (2004). 
        http://www.iop.org/EJ/abstract/0305-4470/37/3/L01/ )     
\end{abstract}


\maketitle

Let us consider an $n$ component network whose dynamics is described by a set of stochastic differential equations \cite{vankampen}:
\begin{equation}
  \dot{q}_{tj} = f_j( {\bf q}_t ) + \zeta_j( {\bf q}_t , t) \, .
\end{equation}
The question is whether or not we can find a potential function from 
Eq.(1) which gives a global description of dynamics. Here  $\dot{q}_{tj} = d q_{tj}/d t$ with $j = 1, 2, ..., n $ and the subscript $t$ for ${\bf q}$ indicates that the state variable ${\bf q}$ is a function of time. The value of $j^{th}$ component is denoted by 
$q_j$. The network state variable forms an $n$ dimensional vector 
${\bf  q}^{\tau} = (q_1, q_2, ... , q_n)$ in the state space. 
Here the superscript $\tau$ denotes the transpose. 
The state variable may be the values of particle coordinates in physics or the protein numbers in a signal transduction pathway, or any other possible quantities specifying the network.
Let $f_j({\bf q})$ be the deterministic nonlinear force on the $j^{th}$ component, which includes both the effects from other components and itself, and $\zeta_j( {\bf q}, t)$ the random force. For simplicity we will assume that $f_j$ is a smooth function explicitly independent of time. 
To be specific, the noise will be assumed to be Gaussian and white with the variance, 
\begin{equation}
  \langle \zeta_i ( {\bf q}_t , t) \zeta_j ({\bf q}_{t'}, t') \rangle 
  = 2 D_{ij}( {\bf q}_t ) \delta (t-t') ,
\end{equation}
and zero mean, $\langle \zeta_j \rangle = 0$. Here $\delta(t)$ is the 
Dirac delta function, and $\langle ... \rangle $ indicates the average with respect to the dynamics of the stochastic force. 
By the physics and chemistry convention the semi-positive definite symmetric matrix $D=\{D_{ij}\}$ with $i,j=1,2, ..., n$ is the diffusion matrix. Eq.(2) also implies that, in situations where the temperature $T$ can be defined, we have set $k_B T = 1$ with $k_B$ the Boltzmann constant. We remark that if an average over the stochastic force ${\bf \zeta}$, a Wiener noise, is performed, Eq.(1) is reduced to the following equation in dynamical systems: 
\[
  \langle \dot{\bf q}_t \rangle 
         = \langle {\bf f}({\bf q}_t ) \rangle 
         = {\bf f}( \langle{\bf q}_t \rangle ) \;  .
\]  
The last equality is due to the fact that at same time $t$, the noise and the state variables are independent of each other. It is equivalent to the fact that the noise can be switch out without affecting the deterministic force, a process demonstrated possible in physics in dealing with environmental effects \cite{leggett}.
A broad range of phenomena in both natural and social sciences has been described by such a deterministic equation \cite{kaplan}. 

Because of its importance and usefulness, repeated attempts have been made to construct a potential function \cite{nicolis,haken,graham,ross}. 
The effort had, however, only limited success\cite{cross}. The usefulness of a potential reemerges in the current study of dynamics of gene regulatory networks \cite{wolynes,zhu}, which would again require its construction in complex network dynamics.
It has been observed that the nonlinear dynamics is in general dissipative (${\rm tr}(F) \neq 0$), asymmetric ($F_{ji} \neq F_{ij}$), and stochastic ($\zeta \neq 0$). 
Here the force matrix $F$ is defined as
\begin{equation}
  F_{ij} = \partial f_i /\partial q_j   \; \;  i,j=1,...,n  \;.
\end{equation}
and the trace is equal to the divergence of the force: ${\rm tr}(F) = \partial \cdot 
{\bf f} = \sum_{j=1}^{n} \partial f_j /\partial q_j  $. 
The combination of those three features prevents any direct application of the insight from Hamiltonian dynamics and has been the main obstacle preventing the potential construction.  In fact, the asymmetry of dynamics has been characterized as the hallmark of the network in a state far from thermal equilibrium, and has been proclaimed that it makes the usual theoretical approach near thermal equilibrium unworkable \cite{nicolis}. 
It is the goal of this paper to report that we have, nevertheless, discovered a novel construction that can take care of those dynamical features and can give us a potential function.

We state, the explicit construction will be given below, that there exists a unique decomposition such that Eq.(1) can be rewritten in the following form:
\begin{equation}
  [ S({\bf q}_t) + A ({\bf q}_t) ] \dot{\bf q}_t 
     = - \partial \phi ({\bf q}_t) + {\bf \xi}({\bf q}_t, t) \; ,
\end{equation}
with the semi-positive definite symmetric matrix $S({\bf q}_t)$, the anti-symmetric matrix $A({\bf q}_t)$, the single-valued scalar function $\phi({\bf q}_t)$, and the stochastic force ${\bf \xi}({\bf q}_t, t)$. 
Here $\partial$ is the gradient operator in the state variable space.
It is straightforward to verify that the semi-positive definite symmetric matrix term is `dissipative': $ \dot{ \bf q}^{\tau}_t S({\bf q}_t) \dot{\bf q}_t \geq 0 $; 
the anti-symmetric part does no `work': 
$ \dot{\bf q}^{\tau}_t A ({\bf q}_t) \dot{\bf q}_t = 0 $, therefore non-dissipative. 
Hence, it is natural to identify that the dissipation is represented by the semi-positive definite symmetric matrix  $S({\bf q})$, the friction matrix, and the transverse force by the anti-symmetric matrix $A({\bf q})$, the transverse matrix. The scalar function 
$\phi({\bf q})$ then acquires the meaning of potential energy. 

The decomposition from Eq.(1) to (4) may be called the $\phi$-decomposition.
However, without further constraint, Eq.(4) would be not unique. 
This may be illustrated by a simple counting: 
There are four apparent independent quantities in Eq.(4), while there are only two in Eq.(1).  
In order to have a unique form for Eq.(4), we may choose to impose the constraint on the stochastic force and the semi-positive definite symmetric matrix in the following manner:
\begin{equation}
 \langle {\bf \xi}({\bf q}_t, t) {\bf \xi}^{\tau} ({\bf q}_{t'}, t') \rangle 
    = 2 S ({\bf q}_t) \delta(t-t') \; ,
\end{equation}
and $\langle {\bf \xi}({\bf q}_t, t) \rangle = 0$.
We observe that this constraint is consistent with the Gaussian and white noise assumption for ${\bf \zeta}$ in Eq.(1). It may be called the stochasticity-dissipation relation. We further observe that that the forms of Eq.(4) and (5) strongly resemble those of dissipative dynamics in quantum mechanics when both dissipative and Berry phase exists \cite{at,leggett}. The constrained $\phi$- decomposition will be called the gauged $\phi$-decomposition, which is indeed unique, as we will now demonstrate. 

We prove the existence and uniqueness of the gauged $\phi$- decomposition from Eq.(1) to (4) by an explicit construction. 
Using Eq.(1) to eliminate the velocity $\dot{\bf q}_t $ in Eq.(4), we have 
\[
   [S({\bf q}_t) + A({\bf q}_t)] [ {\bf f}({\bf q}_t)
                 + {\bf \zeta}({\bf q}_t,t) ]
     = - \partial \phi({\bf q}_t)+ {\bf \xi}({\bf q}_t,t) \; .
\]
Noticing that the dynamics of noise is independent of that of the state variables we require that both the deterministic force and the noise satisfying following two equations separately. For the deterministic force, this leads to 
\begin{equation}
  [S({\bf q}) + A({\bf q})] {\bf f}({\bf q}) 
   = - \partial \phi({\bf q}) \; ,
\end{equation}
suggesting a `rotation' from the force ${\bf f}$ to the gradient of the potential $\phi$ at each  point in the state space. We have dropped the subscript $t$.
For stochastic force, we have: 
\begin{equation}
  [S({\bf q}) + A({\bf q})] {\bf \zeta}({\bf q},t) 
   =  {\bf \xi} ({\bf q},t) \; ,
\end{equation}
which shows the same `rotation'  between the stochastic forces. Here we have also dropped the subscript $t$ for the state variable. 
Using Eq.(2) and (5), Eq.(7) implies 
 \begin{equation}
   [S({\bf q}) + A({\bf q}) ] D({\bf q}) [S({\bf q}) - A({\bf q}) ] 
     = S({\bf q}) \; ,
\end{equation}
which suggests a duality between Eq.(1) and (4): a large friction 
matrix implies a small diffusion matrix. It is a generalization of the 
Einstein relation \cite{einstein} to finite transverse matrix $A$.

Next we introduce an auxiliary matrix function 
\begin{equation}
   G({\bf q}) = [S({\bf q}) + A({\bf q}) ]^{-1} \; .
\end{equation}
Here the inversion `${-1}$' is with respect to the matrix. Using the 
property of the potential function $\phi$: 
$\partial \times \partial \phi = 0$, 
Eq.(6) leads to
\begin{equation}
  \partial \times [ G^{-1} {\bf f}({\bf q}) ] = 0 \; ,
\end{equation}
which gives $n (n-1)/2$ conditions to determine the $n\times n$ auxiliary matrix $G$.
The generalized Einstein relation, Eq.(8), leads to the following 
equation
\begin{equation}
   G + G^{\tau} = 2 D \; ,
\end{equation}
which readily determines the symmetric part of the auxiliary matrix 
$G$, another $n (n+1)/2$ conditions for $G$. Eq. (10) and (11) give the needed $n\times n$ conditions to completely determine $G$. Here we give a solution of $G$ as a series in gradient expansion:
\begin{equation}
   G = D + \sum_{j=0}^{\infty} \Delta G_j \; , 
\end{equation}
with $\Delta G_j = \sum_{l=1}^{\infty} (-1)^l [ (F^{\tau})^l  \tilde{D}_j F^{-l} + (F^{\tau})^{-l} \tilde{D}_j F^l ] $,  
$\tilde{D}_0 = DF - F^{\tau}D$,  
$\tilde{D}_{j \geq 1} = ( D + \Delta G_{j-1} ) 
         \left\{ [\partial \times (D^{-1} + \Delta G_{j-1}^{-1} ) ] 
         {\bf f} \right\} ( D - \Delta G_{j-1} )$. 
The zeroth order approximation to Eq.(10) is $G F^{\tau} - F G^{\tau} = 0$. A formal solution to this approximated equation together with Eq.(11) has been constructed under a rather restrictive condition \cite{ao}, and the explicit solution under a generic condition has been obtained in Ref. [14].
The gauged $\phi$-decomposition is therefore uniquely determined: 
\begin{equation}
  \left\{ \begin{array}{lll}
  \phi({\bf q}) & = & - \int_C d{\bf q}' \cdot 
          [ G^{-1}({\bf q}') {\bf f}({\bf q}') ] \\
  S({\bf q})    & = & [G^{-1}({\bf q}) 
          + (G^{\tau} )^{-1}({\bf q}) ] /2 \\
  A({\bf q})   & = & [G^{-1}({\bf q}) - (G^{\tau} )^{-1}({\bf q}) ] /2
  \end{array} 
     \right. \; . 
\end{equation} 
The end and initial points of the integration contour $C$ are $\bf q$ and ${\bf q}_0$ respectively.
During the construction a sufficient condition $\det(F) \det(S+A)\neq 
0$ is assumed, with exception at some isolated points.
We remark on the special role played by the force matrix: If $F D = D 
F^{\tau}$, i.e., $\tilde{D}_0 = 0$. If in addition $\tilde{D}_{j \geq 1}=0$ in this case, which can be realized if $D$ is a constant, 
$\Delta G_j = 0$ for all $j$. This means that $G = D$ and the transverse matrix $ A=0$. Such a condition has been noticed in the linear case where both $F$ and $D$ are constant matrices, and named the integrability condition \cite{lax}.  

In many experimental studies of a complex network, a question is often asked on the distribution of the state variable after a transient period instead of focusing on the individual trajectory of the network. 
This implies that either there is an ensemble of identical networks or repetitive experiments are been carried out. From statistical mechanics, if viewing the potential function $\phi$ as an energy, a steady distribution function can be expected from Eq.(4):
\begin{equation}
   \rho_0({\bf q}) = \frac{1}{Z} \; \exp\{ - \phi({\bf q}) \} \; ,
\end{equation}
with the partition function $Z = \int d^n {\bf q} \; \exp\{ - \phi({\bf  q}) \} $.
This is a Boltzmann-Gibbs distribution for the state variable, and give the strongest manifestation of the usefulness of the potential function $\phi$. 
We remark that it is however not obvious that the steady state distribution, if exists, should be given by Eq.(14). In the following we give a heuristic demonstration that Eq.(14) is indeed a right steady distribution for the network as the steady state solution of the corresponding Fokker-Planck equation. 

The connection between the standard stochastic differential equation, Eq.(1), and 
Fokker-Planck equation is necessarily ambiguous for the generic nonlinear case as exemplified by the Ito-Stratonovich dilemma \cite{vankampen,gardiner,risken}. 
We attribute this lack of definiteness to the asymptotic nature of the connection in which a procedure must be explicitly defined: Different procedures will in general lead to different results. Here we present another procedure which may be natural from the theoretical physics point of view. Our starting point will be Eq.(4),  not Eq.(1) from which most previous derivations started.

The existence of both the deterministic force and the stochastic force in Eq.(4) suggests that there are two well separated time scales in the network: microscopic time scale for the description of the stochastic force and macroscopic time scale for the network motion. The former time scale is much smaller than the latter. This separation of time scales further suggests that the macroscopic motion of the network has an inertial: It cannot response instantaneously to the microscopic motion. 
To capture this feature, we introduce a small inertial "mass" $m$ and a kinetic momentum vector ${\bf p}$ for the network. The dynamical equation for the network now takes the form:
\begin{equation}
   \dot{\bf q}_t = {\bf p}_t /m 
\end{equation}
defines the kinetic momentum, and 
\begin{equation}
    \dot{\bf p}_t = - [S({\bf q}_t) + A({\bf q}_t) ] {\bf p}_t /m 
                         - \partial \phi({\bf q}_t) 
                         + {\bf \xi}({\bf q}_t, t)
\end{equation} 
is the extension of Eq.(4). We note that there is no dependent of friction matrix and the stochastic force on the kinetic momentum, therefore no Ito-Stratonovich dilemma in the connection between the stochastic differential equation and the dynamical equation for the distribution function. The Fokker-Planck equation in this enlarged state space, the 
Klein-Kramers equation, can be immediately obtained \cite{vankampen}:
\begin{equation}
  \left\{ \partial_t + \frac{\bf p}{m} \cdot \partial_{\bf q} 
           + \overline{\bf f} \cdot \partial_{\bf p} 
  - \partial_{\bf p} S \left[\frac{\bf p}{m} + \partial_{\bf p} \right]
\right\} 
   \rho({\bf q}, {\bf p}, t) = 0 \; . 
\end{equation}
Here $\overline{\bf f} = {\bf p} A /m - \partial_{\bf q} \phi $, and $t$, ${\bf q}$, and ${\bf p}$ are independent variables. The steady  distribution can be found as \cite{vankampen} 
\begin{equation}
 \rho({\bf q},{\bf p}) 
        = \exp\{ - [ {\bf p}^2 /2m + \phi({\bf q}) ] \} / Z \; ,
\end{equation}
with $Z = \int d^n{\bf q} d^n {\bf p} \exp\{ - [ {\bf p}^2 /2m + \phi({\bf q}) ] \}$ the partition function. 
There is an explicit separation of state variable and its kinetic momentum in Eq.(18). The zero "mass" limit can then be taken with no effect on the state variable distribution. This confirms that Eq.(14), the Boltzmann-Gibbs distribution expected from Eq.(4), is the right choice under this procedure. 

To conclude, we point out a major difference between the present construction of the potential and those in literature such as represented by the Graham-Haken construction \cite{graham,haken}: The present construction is based on a structure built into stochastic differential equation. There is no explicit use of Fokker-Planck equation. Therefore there is no need to make assumption on the distribution function in the limit time goes to infinite as assumed in the Graham-Haken construction. In particular, the potential in the present paper can be time dependent.

Critical discussions with N. Goldenfeld, A.J. Leggett, C. Kwon, J. Ross, D.J. Thouless, 
L. Yin, and X.-M. Zhu are highly appreciated. This work was supported in part by a USA NIH grant under the number HG002894-01.

\end{document}